\documentclass{elsart}
\usepackage{graphics}
\usepackage{graphicx}
\usepackage{epsfig}

\begin{document}

\begin{frontmatter}



\title{Universal properties of population dynamics with fluctuating resources
\thanksref{talk}}
\thanks[talk]{Expanded version of a poster presented at the Stat-Phys 2007
 meeting on Statistical Physics (Kolkata, January 2007).}


\author{Sayak Mukherjee$^1$, H.K. Janssen$^2$, 
and B. Schmittmann$^1$\corauthref{cor}}
\ead{smukhe04@vt.edu, janssen@thphy.uni-duesseldorf.de, and schmittm@vt.edu }
\corauth[cor]{Corresponding author}
\address{$^1$Department of Physics, Virginia Tech, Blacksburg, VA 24061-0435, USA\\
$^2$ Institut f\"{u}r Theoretische Physik III, Heinrich-Heine-Universit\"{a}t,\\
40225 D\"{u}sseldorf, Germany.}

\begin{abstract}
Starting from the well-known field theory for directed percolation, we describe 
an evolving population, near extinction, in an environment with its own nontrivial 
spatio-temporal dynamics. Here, we consider the special case 
where the environment follows a simple relaxational (Model A) dynamics. 
Two new operators emerge, with upper critical dimension of four, which couple 
the two theories in a nontrivial way. While the Wilson-Fisher fixed point 
remains completely unaffected, a mismatch of time scales destabilizes the usual 
DP fixed point, suggesting a crossover to a first order transition from 
the active (surviving) to the inactive (extinct) state. 
\end{abstract}

\end{frontmatter}


\emph{Introduction.}\label{sec1} The understanding of ecological
catastrophies, where one or several species become extinct, is a major
challenge in various areas of science. From a physics perspective,
extinction events are frequently associated with a continuous phase
transition from an active to an inactive (absorbing) state from which the
population cannot recover. As the transition is approached, the dynamics is
characterized by large fluctuations, whose characteristic length scale
diverges. By virtue of this diverging scale, the large-distance long-time
behavior of systems near continuous phase transitions is universal, i.e.,
independent of microscopic detail, and can be captured successfully by
minimal models. Renormalization group (RG) approaches have been extremely
successful in exploring the physics of such models near criticality. In this
brief note, we explore the effects of a fluctuating environment, reflecting,
e.g., a food supply with its own nontrivial dynamics, on the universal
properties of this active-inactive state transition. We first summarize the
field theory for this transition in a uniform environment. Next, we
introduce the dynamics of the environment and discuss its effects on the
evolving population. We conclude with a brief summary and some open
questions.

\emph{The model. }The minimal model describing the extinction of a single
species in a uniform environment is well known \cite{cardy,janssen,hk}. It finds
its applications in a broad range of problems, ranging from population
dynamics to catalysis, forest fires, and directed percolation (DP). It is
most easily expressed in the language of ``chemical'' reactions, describing
the birth ($A\rightarrow 2A$), death ($A\rightarrow \emptyset $), and
overcrowding ($A+A\rightarrow A$) of a population, with rates $\sigma $, $%
\mu $, and $g/2$, respectively. The individuals are free to diffuse in $d$
dimensions. Once the last individual has died, the species cannot recover.
Denoting the space- and time-dependent density of the population by $s(x,t)$%
, the associated field theory is well known \cite{cardy,janssen} and can be written
in terms of a dynamic functional, with a response field $\tilde{s}(x,t)$:\ 
\begin{equation}
J_{DP}\{\tilde{s},s\}=\int d^{d}x\int dt\,\lambda \tilde{s}\{\lambda
^{-1}\partial _{t}+\left( \tau -\bigtriangledown ^{2}\right) +\frac{1}{2}%
\left( gs-\bar{g}\tilde{s}\right) \}s
\end{equation}%
Here, $\tau \equiv \mu -\sigma $ is the net death rate which controls the
distance from the (mean-field) critical point, and $\lambda $ sets the time
scale. The stochastic character of the dynamics, along with the absorbing
state condition, is reflected in the noise vertex, $\bar{g}\tilde{s}^{2}s$.
Invariance under rapidity reversal implies $g=\bar{g}$. For now, we retain
different nonlinearities here, to allow for a breaking of this symmetry in
the full model. The effects of the environment are encoded in the \emph{%
constant }rates $\tau $, $g$, and the time scale $\lambda $. It is natural
to study the consequences of fluctuating rates, reflecting an environment
that has its own nontrivial spatio-temporal structure. Previously, such
effects as a quenched random $\tau (x)$ \cite{qr} or a coupling to a
diffusive mode \cite{schmittmann,vW,Oe} were investigated. Here, we study an environment
with non-conserved relaxational dynamics, described by Model A \cite{hohenberg,wh}.
Denoting the Model A local order parameter as $\varphi $, modelling, e.g., a
nutrient supply for the population, the corresponding field theory is given
by  
\begin{equation}
J_{A}\{\tilde{\phi},\phi \}=\int d^{d}x\int dt\gamma \{\tilde{\varphi}\left[
\gamma ^{-1}\partial _{t}\varphi +\left( r-\bigtriangledown ^{2}\right)
\varphi +\frac{u}{3!}\varphi ^{3}\right] -\tilde{\varphi}^{2}\}
\end{equation}%
where $\tilde{\varphi}$ is, again, the corresponding response field, and $r$
is the critical parameter. Noting that both field theories have an upper
critical dimension $d_{c}=4$, we seek novel couplings, involving both Model
A and DP fields, which might destabilize the familiar DP \cite{janssen} or
Wilson-Fisher \cite{WFFP} fixed points. If $\mu $ is an external momentum
scale, naive dimensional analysis results in $r\sim \tau \sim \mu ^{2}$.
Assuming that the net death rate depends on the environment (e.g., via the
availability of nutrients) and that the presence of the population affects
its environment, it is natural to write an expansion $\tau =\tau _{o}+\kappa
\varphi +w\varphi ^{2}+...$ and $r=r_{o}+vs+...$. Dimensional analysis shows
that $\kappa \sim \mu ^{(6-d)/2}$, $v\sim \mu ^{(4-d)/2}$, and $w\sim \mu
^{4-d}$. The coupling $\kappa $ is relevant in $d=4$ and must be tuned to
zero to access the multi-critical point $r=\tau =\kappa =0$. Alternately, if
we demand that the up-down symmetry of the Ising model remain valid, $\kappa 
$ can be set to zero for physical reasons. Once zero, it is not generated
under the RG. Collecting, our model is given by $J\{\tilde{\phi},\phi ,%
\tilde{s},s\}=J_{DP}\{\tilde{s},s\}+J_{A}\{\tilde{\phi},\phi \}+J_{int}\{%
\tilde{\phi},\phi ,\tilde{s},s\}$ where 
\begin{equation}
J_{int}\{\tilde{\phi},\phi ,\tilde{s},s\}\equiv  \int d^{d}x\int dt\left\{
\gamma v\tilde{\phi}\phi s+\frac{1}{2}\lambda w\tilde{s}s\phi ^{2}\right\} 
\end{equation}%
Expectation values of the four fields are given as functional integrals with
weight $\exp \left( -J\right) $.

\emph{RG results. }All five nonlinear couplings - $u$, $g$, $\bar{g}$, $v$,
and $w$ - are marginal in the upper critical dimension $d_{c}=4$. The new
vertex $v$ violates the invariance under rapidity reversal but respects the
absorbing state condition. We use renormalized perturbation theory, in $%
\epsilon =d_{c}-d$, combined with minimal subtraction. Due to the absence of
a bare correlator in the DP field theory, no corrections to the Model A
correlation and response functions are generated. As a result, the Model A
fixed point remains at the Wilson-Fisher value, and all Model A exponents
retain their familiar values. There are, however, corrections to the DP
couplings, and nontrivial flow equations for the new mixed couplings. In
one-loop order, we find numerous infrared unstable fixed points, and one
nontrivial stable one, characterized by the \emph{renormalized} values $%
u=2\epsilon /3+O(\epsilon ^{2})$, $g\bar{g}=4\epsilon /3+O(\epsilon ^{2})$, $%
v\bar{g}=0+O(\epsilon ^{2})$, and $%
w=13\left( \lambda +\gamma \right) /\left( 12\lambda \right) $. A geometric
factor $G_{\epsilon }\equiv 2\Gamma \left( 1+\epsilon /2\right) \left( 4\pi
\right) ^{-d/2}$ occurs in each loop integral and has been absorbed in the
definition of the couplings. At first glance, this looks encouraging; in
particular, $v\bar{g}=0$ restores the rapidity reversal symmetry. However,
at this fixed point, the ratio $\rho \equiv \gamma /\lambda $ of the two
time scales flows towards $\rho =0$. This implies that the dynamics of Model
A freezes on the time scale of the DP fields. The limit $\rho \rightarrow 0$
turns out to be singular, so that the theory needs to be be reanalyzed. At the most naive level, this involves
adding the time-delocalized vertex, $\bar{w}\int d^{d}x\int dt\int
dt^{\prime }\,\tilde{s}(x,t)s(x,t)\tilde{s}(x,t^{\prime })s(x,t^{\prime })$,
to the DP field theory. Once included, however, \emph{all} fixed points
become unstable \cite{hk,qr}, indicating that the active-inactive state
transition might turn first order. At a more rigorous level, one should
integrate out the Model A fields and then take the static limit. The
resulting field theory will be analyzed elsewhere \cite{JMS-new}.

\emph{Conclusions. }In summary, we have investigated the effect of a
fluctuating environment on the evolution of a population near the brink of
extinction. These fluctuations may be due to a food source, for example,
which has its own nontrivial spatio-temporal dynamics. Here, we consider the
interactions of the well-known DP\ field theory with an auxiliary field
obeying simple relaxational (Model A) dynamics. Investigating the combined
field theory to one-loop order in $\epsilon =4-d$, we find that the RG\ flow
implies a freezing of the Model A dynamics on the times scales of the DP
fields. Reanalyzing the theory in this (singular) limit, all fixed points
are now infrared unstable, indicating a possible first order transition. Two
open questions remain, namely, first, to quantify the crossover to the
time-delocalized theory, and second, to compare these field-theoretic
findings near $d_{c}=4$ to Monte Carlo simulations in two and three
dimensions.

\emph{Acknowledgements. } We thank U.C. T\"{a}uber and R.K.P. Zia for helpful
discussions. This work is supported in part by the NSF through DMR-0414122 and 
SBE-0244916.

\bigskip 


\end{document}